\documentclass[%
 reprint,
 superscriptaddress,
 groupedaddress,
 amsmath,amssymb,
 aps,
 prl,
]{revtex4-1}

\usepackage{amsmath}
\usepackage{graphicx}  
\usepackage{amssymb}   
\usepackage{hyperref}
\usepackage{color}
\hypersetup{
    pdftitle={article},    
    pdfauthor={Siddharth Paliwal},     
    pdfsubject={Active Systems},   
    colorlinks=true,       
    linkcolor=red,          
    citecolor=blue,        
    final=true
}

\begin{document}

\title{The role of topological defects in the two-stage melting and \\ elastic behavior of active Brownian particles}

\author{Siddharth Paliwal}
\email{s.paliwal@uu.nl}
\author{Marjolein Dijkstra}
\email{m.dijkstra@uu.nl}
\affiliation{Soft Condensed Matter, Debye Institute for Nanomaterials Science, Utrecht University, Princetonplein 1, 3584 CC Utrecht, The Netherlands}

\date{\today}

\renewcommand{\sectionautorefname}{Sec.}			
\renewcommand{\subsectionautorefname}{Sec.}		
\renewcommand{\equationautorefname}{Eq.}
\renewcommand{\figureautorefname}{Fig.}

\begin{abstract}
We find  that crystalline states of repulsive active Brownian particles at high activity melt into a hexatic phase but this transition is not driven by an unbinding of bound dislocation pairs as suggested by the Kosterlitz-Thouless-Halperin-Nelson-Young (KTHNY) theory. Upon reducing the density, the crystalline state  melts into a high-density hexatic state  devoid of any defects. Decreasing the density further, the dislocations proliferate and introduce plasticity in the system, nevertheless maintaining the hexatic state, but eventually melting into a fluid state. Remarkably, the elastic constants of active solids  are equal to those of their passive counterparts, as the swim contribution to the stress tensor is negligible in the solid state. The sole effect of activity is that the stable solid regime shifts to higher densities.
Furthermore, discontinuities in the elastic constants as a function of density correspond to changes in the defect concentrations rather than to the solid-hexatic transition.
\end{abstract}

\maketitle

According to the Kosterlitz-Thouless-Halperin-Nelson-Young (KTHNY) theory, a two-dimensional solid of passive particles  melts via a continuous transition into an intermediate hexatic state of quasi-long-ranged bond orientational order, and melts subsequently via a second continuous transition into a fluid state~\cite{Kosterlitz1973,Nelson1979,Chui1982}. These transitions are triggered by the unbinding of topological defects, which are particles with a non-conforming number of neighbors with respect to that of the crystal lattice. The coordination number is 6 for  an ideal triangular lattice. Hence, particles with a number of neighbors $N_b$ that deviates from 6 are classified as defects. One can distinguish defects that either exist freely as 5-fold and 7-fold disclinations or cluster into dislocations (5-7 pairs), dislocation pairs (5-7-5-7 quartets) or higher-order clusters.  
The debate on the melting behavior of an equilibrium system of hard disks as well as short-ranged repulsive disks was only settled a few years ago, both in simulations~\cite{Bernard2011, Engel2013, Qi2014, Kapfer2015, anderson2017shape}  and experiments~\cite{Thorneywork2017}, showing a two-stage melting scenario, that deviates from the KTHNY scenario, of  a continuous solid-hexatic transition driven by an unbinding of dislocation pairs and a first-order hexatic-fluid transition due to a proliferation of grain boundaries.

	\begin{figure}
	\centering
	\includegraphics[width=\columnwidth]{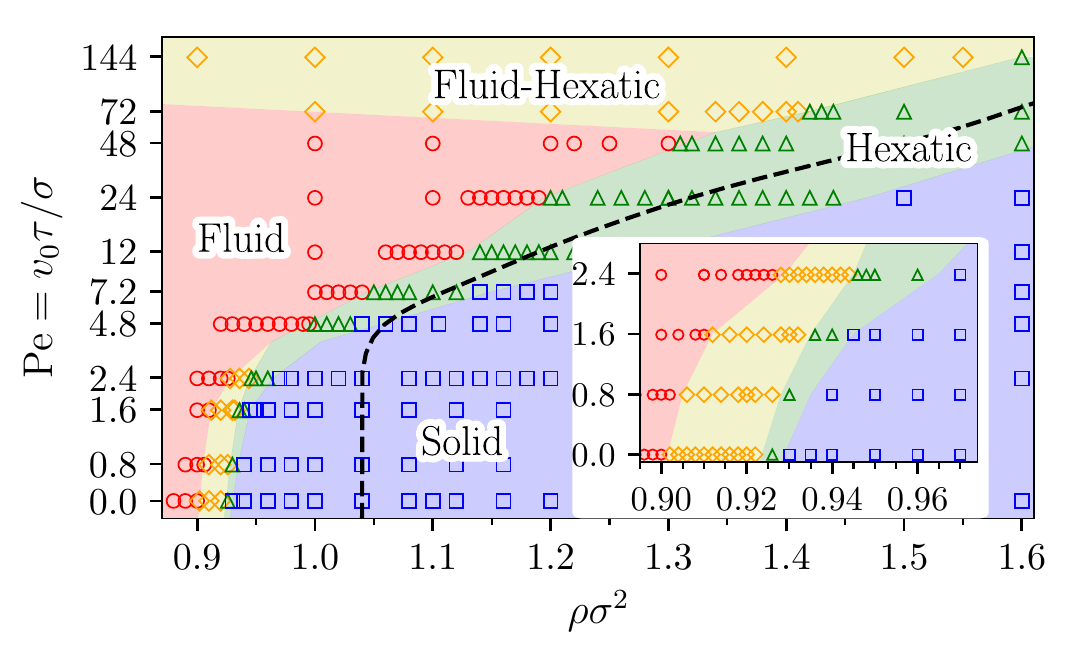}
	\caption{State diagram in the Pe-$\rho\sigma^2$ representation exhibiting  fluid (circle, red), fluid-hexatic coexistence (diamonds, yellow), hexatic (triangle, green) and crystal (square, blue) states. The symbols denote  state points used in the simulations, the background colors denote the boundaries of the labeled regions, and the black dashed line indicates the densities beyond which the concentration of topological defects vanishes in simulations of $N=72\times10^3$ particles. The inset is a magnification of the low-Pe regime.
	\label{fig:statedia}}
	\end{figure}	
	
Remarkably, non-equilibrium systems of self-propelled particles~\cite{Fily2012,Redner2013,Stenhammar2014,Siebert2018}, which constantly convert energy from the environment into persistent motion, have recently also been shown to follow such a two-stage melting behaviour~\cite{Digregorio2018,Klamser2018}. It was found that the first-order nature of the liquid-hexatic transition persists upto a small degree of activity. The transition then becomes continuous until it reappears as a coexistence of dilute and dense states at high activity~\cite{Digregorio2018,Klamser2018}. The solid state was found to melt into a hexatic state via a continuous transition. In this work, we further explore this hexatic-solid transition with respect to the role of topological defects discussed in the KTHNY theory and investigate whether the melting is driven by transitions in defect concentrations.

	\begin{figure*}
	\centering	
		\includegraphics[width=\textwidth]{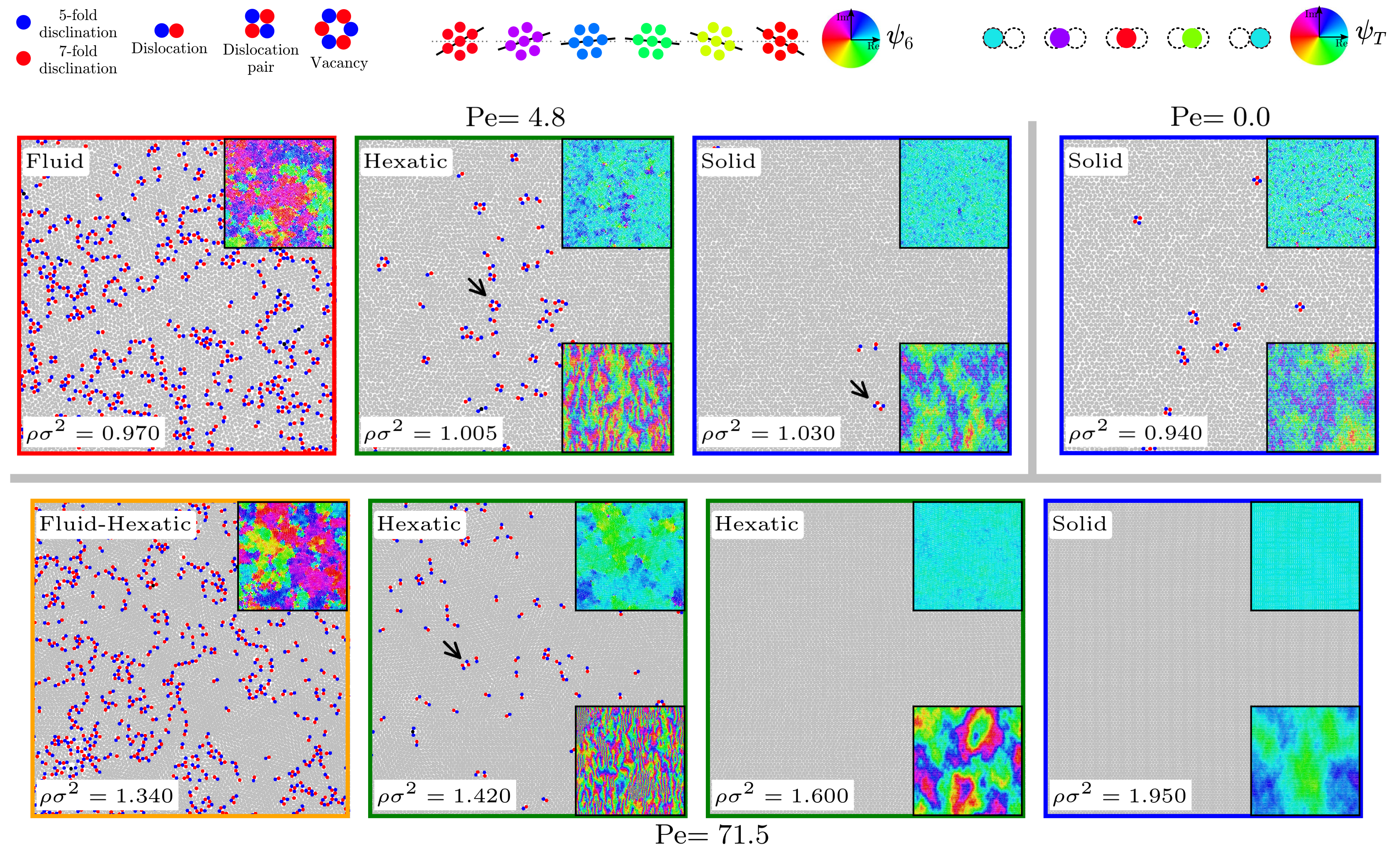}
	\caption{Typical sections of active Brownian particle configurations ($80\sigma\times80\sigma$) for $\mathrm{Pe}=0.0, 4.8$ and $71.5$ at labeled states, showing   5-fold (blue) and 7-fold (red) defects, other defects (black), and particles with $N_b=6$ (grey). Some vacancies are indicated by arrows in the hexatic and solid states. The configurations at $\mathrm{Pe}=4.8$ are similar to the corresponding passive states. For $\mathrm{Pe}=71.5$ we find that with increasing density the hexatic states show a decrease in the number fraction of defects, with a complete absence of defects at a density $\rho\sigma^2\gtrsim1.560$. The positional correlations become quasi-long-ranged around a density $\rho\sigma^2\simeq1.950$ for $\mathrm{Pe}=71.5$ as shown in \autoref{fig:statedia}. The insets in the upper right and lower right show the same configurations as the main panels but colored according to the hexatic and positional order parameters $\psi_{6}$ and $\psi_{T}$, respectively, following the color mapping shown at the top.}
	\label{fig:snapdefects}
	\end{figure*}
 
We numerically simulate a system of $N$ active Brownian particles exhibiting overdamped Langevin dynamics:
\[ \gamma\dot{{\mathbf{r}}}_i = -\sum_{j \neq i}^{} \boldsymbol{\nabla}_i \mathrm{U} (r_{ij}) + \gamma v_0\mathbf{e}_i	+ \sqrt{2\gamma k_BT} \boldsymbol{\Lambda}_i^{t},
\]
where $v_0\mathbf{e}_i$ is the self-propulsion speed, $\gamma$ is the damping coefficient, $k_B$  the Boltzmann constant, and $T$  the temperature of the solvent. The particle orientation $\mathbf{e}_i=(\cos\theta,\sin\theta)$ undergoes free rotational diffusion $\dot{\theta}_i = \sqrt{2D_r}\Lambda_i^r$, where $D_r$ is the rotational diffusion coefficient. The quantities $\boldsymbol{\Lambda}_i^{t}$ and $\Lambda_i^{r}$ are unit-variance Gaussian noise terms with zero mean. The particles interact with a pairwise repulsive WCA potential $\mathrm{U} (r) = 4\varepsilon\left[ \left(\frac{\sigma}{r}\right)^{12} -\left(\frac{\sigma}{r}\right)^6 \right] + \varepsilon$, and we set $k_BT/\varepsilon = 1$. We first determine the phase boundaries by measuring the equation of state (pressure-density curves), density histograms,  and the decay of the orientational and  positional correlation functions. We present the state diagram in \autoref{fig:statedia} in the activity-density (Pe-$\rho$) representation, where the P\'eclet number is defined by $\mathrm{Pe}=v_0\tau/\sigma$. The state diagram  shows that  the continuous hexatic-solid transition persists upon introducing activity, but shifts to higher densities due to the softness of the interaction potential. We hereby assume that the solid state transforms into a hexatic phase when the positional correlations decay with a power law exponent $\eta_T\le1/3$ as described by the equilibrium KTHNY theory.

\emph{Topological defects:\label{sec:defects}}
We identify the topological defects by performing a Voronoi construction and calculating the number of neighbors $N_b$ for each particle. We classify them as $N_b$-fold defects and  account for 5-fold and 7-fold defects.  
In \autoref{fig:snapdefects} we show typical configurations, highlighting these defects in the fluid, hexatic and  crystalline state at $\mathrm{Pe}=0, 4.8$ and $71.5$. At low activity $\mathrm{Pe}=4.8$, similar defect configurations are observed as for the passive hard-disk systems~\cite{Qi2014}. A finite number fraction of dislocation pairs can be identified in the solid state which can exist due to thermal fluctuations without disturbing the positional order. In the hexatic state, the presence of unpaired dislocation defects causes the positional order to decay exponentially  but the orientational order decays algebraically. Finally the fluid state comprises of many defect clusters and  5-fold and 7-fold disclinations, both destroying  the local bond orientational order. However, the hexatic states at $\mathrm{Pe}=71.5$, are significantly different in defect configurations. The low-density hexatic state at $\rho\sigma^2=1.420$ shows a high number fraction of  dislocations with almost no dislocation pairs, whereas the high-density hexatic state ($\rho\sigma^2=1.600$) at the same Pe shows a complete absence of any defects.

\begin{figure}
	\centering
		\includegraphics[width=\columnwidth]{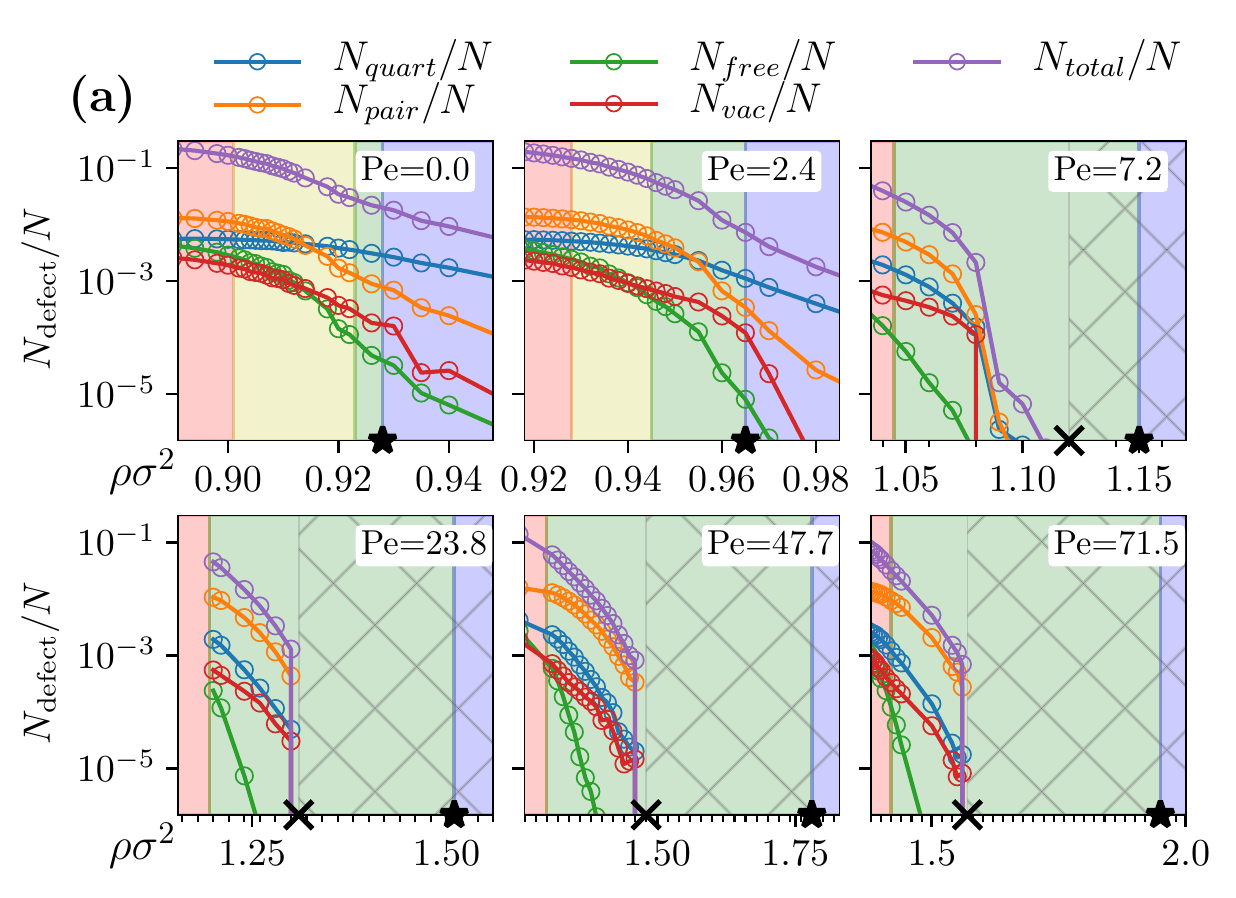}
		\includegraphics[width=0.85\columnwidth]{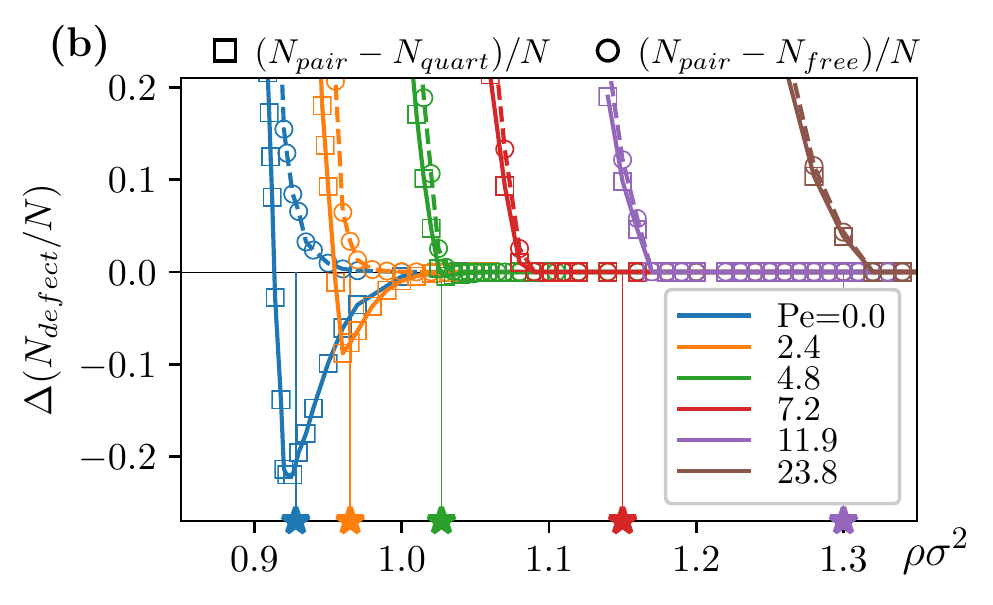}
		\caption{(a) Number fraction of  specific defects as a function of density $\rho\sigma^2$ for various Pe. The background colors mark the boundaries as indicated in the state diagram (\autoref{fig:statedia}) and the region devoid of any defects is cross-hatched. The symbols on the axis indicate the densities where the defects become absent ($\times$), and the hexatic-solid transition densities ($\star$). (b) The difference in the number fraction of unpaired dislocations and bound dislocation pairs $(N_\mathrm{pair}-N_\mathrm{quart})/N$ (squares), and the difference in the number fraction of unpaired dislocations and free disclinations $(N_\mathrm{pair}-N_\mathrm{free})/N$ (circles) as a function of  $\rho\sigma^2$ for various Pe as labeled in the legend.}
	\label{fig:deffract2}
	\end{figure}

We further quantify this observation by measuring the number fractions of the different types of defects, which are plotted in \autoref{fig:deffract2}(a) as a function of density for various Pe. We  clearly observe that the overall number fraction of defects $N_\mathrm{total}/N$ increases upon decreasing the density for all Pe. At low $\mathrm{Pe}\leq 2.4$, we find that the solid phase contains mainly bound dislocation pairs, which  move freely in the solid phase. The number fraction of bound dislocation pairs $N_\mathrm{quart}/N$ increases slightly upon approaching the hexatic-solid transition, whereas the number fraction of dislocations $N_\mathrm{pair}/N$ increases more rapidly. At the hexatic-solid transition, $N_\mathrm{quart}/N$ is still higher than $N_\mathrm{pair}/N$ but at lower density this scenario is reversed. The high fraction of $N_\mathrm{pair}/N$ implies that the quasi-long-range positional order of the system is destroyed, suggesting that the solid-hexatic transition is induced by the unbinding of dislocation pairs.  Upon reducing the density further, we find that the fraction $N_\mathrm{free}/N$ of  5- and 7-fold defects also starts to increase due to the unbinding of dislocations into disclinations and the bond orientational order decays exponentially in the stable liquid phase. To summarize, we find that  the two-step melting scenario consisting of a solid-hexatic transition driven by the unbinding of dislocation pairs and a hexatic-fluid transition caused by defect clusters as observed for the 2D passive systems of short-range repulsive particles, persists at low activity.  

For higher activity $\mathrm{Pe}\geq7.2$, the behavior of dislocations is different than for passive systems as shown in \autoref{fig:snapdefects}. For $\mathrm{Pe}\geq7.2$ the system seems to support a higher fraction of dislocations as compared to dislocation pairs for all densities, as observed in \autoref{fig:deffract2}(a) where $N_\mathrm{quart}$ never exceeds $N_\mathrm{pair}$. This observation is more clearly visible in \autoref{fig:deffract2}(b) where we plot the differences $(N_\mathrm{pair}-N_\mathrm{quart})/N$ and $(N_\mathrm{pair}-N_\mathrm{free})/N$. We find that for $\mathrm{Pe}\leq2.4$ and for high densities $(N_\mathrm{pair}-N_\mathrm{quart})/N$ is negative whereas for $\mathrm{Pe}\geq7.2$ this difference is always positive, thereby demonstrating that  a higher fraction of isolated dislocations over bound ones is favored at higher activity.

In \autoref{fig:deffract2}(a) and (b) we also mark the hexatic-solid transition densities as identified from the spatial decay of positional correlations~\cite{SI}. These densities agree closely with the minimum in $(N_\mathrm{pair}-N_\mathrm{quart})/N$ which corresponds to a reversal in the trend of defect concentrations and confirms our previous observation that for $\mathrm{Pe}\leq 2.4$ the hexatic-solid transition is driven by the unbinding of dislocation pairs. The difference $(N_\mathrm{pair}-N_\mathrm{free})/N$, however, is always positive for all densities and show similar trends for all values of Pe considered.

In \autoref{fig:deffract2}(a) we also locate the densities above which we find a complete absence of defects in the system for all values  of $\mathrm{Pe}$, marked by a cross on the $x$-axis and the corresponding density range by cross-hatching. As we increase the activity, this point crosses over from the dense crystal state to the  hexatic state. For $\mathrm{Pe}=23.8, 47.7$ and $71.5$ we clearly see that a large region corresponding to the dense hexatic states is free from any kinds of topological defects in contrast to the low activity systems. This dense hexatic region, devoid of any defects, still has an exponentially decaying positional order as indicated by the background colors corresponding to the state diagram (\autoref{fig:statedia}). The activity-induced fluctuations decorrelate the particle positions at long range and are responsible for a faster decay of positional order in this density regime. The solid-hexatic transition in active systems is thus driven by a striking non-equilibrium feature. We now test the correspondence of changes in defect concentrations and the elastic response of the system with respect to the predictions of the KTHNY theory.

\emph{Elastic Moduli:\label{sec:elastic}}
The KTHNY theory for the melting of 2D equilibrium solids is based on the linear elastic properties of a continuum. Although the equilibrium theory relies on the elastic deformation energies to resolve the transition to a fluid state with a vanishing shear and renormalized Young's modulus, we can extend the notion of mechanical stress, which is well-defined for an isotropic active fluid, to describe the elastic behaviour in terms of the response to an externally imposed linear strain on the simulation box. In our simulations, we start from a perfect hexagonal initial configuration with $N=2.8\times10^3$ particles and measure the full stress tensor $P_{\alpha\beta}$, comprising of the ideal, virial and swim components~\cite{winkler2015virial}, in the deformed box due to a fixed small linear strain $\epsilon_{xx}\in[-0.01,0.01]$. We calculate the Lam\'e elastic coefficients $\lambda$ and $\mu$ from the effective stiffness tensor $\mathbb{B}$ obtained from the slope of a linear fit to the stress versus strain curves~\cite{Landau1989theory,Ray1988,Frenkelsmit2001} (see Ref.~\cite{SI} for details).

	\begin{figure}
	\centering
	\includegraphics[width=\columnwidth]{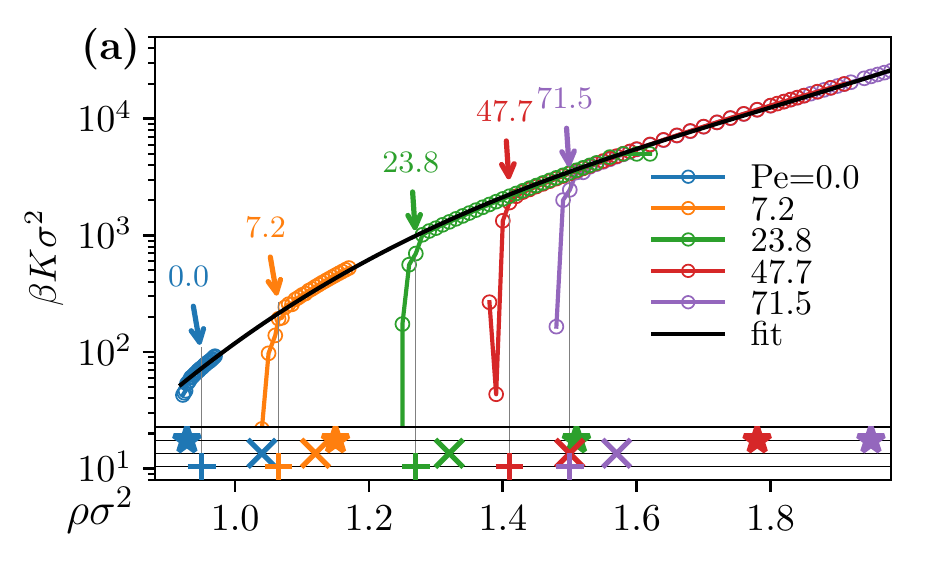}
	\includegraphics[width=\columnwidth]{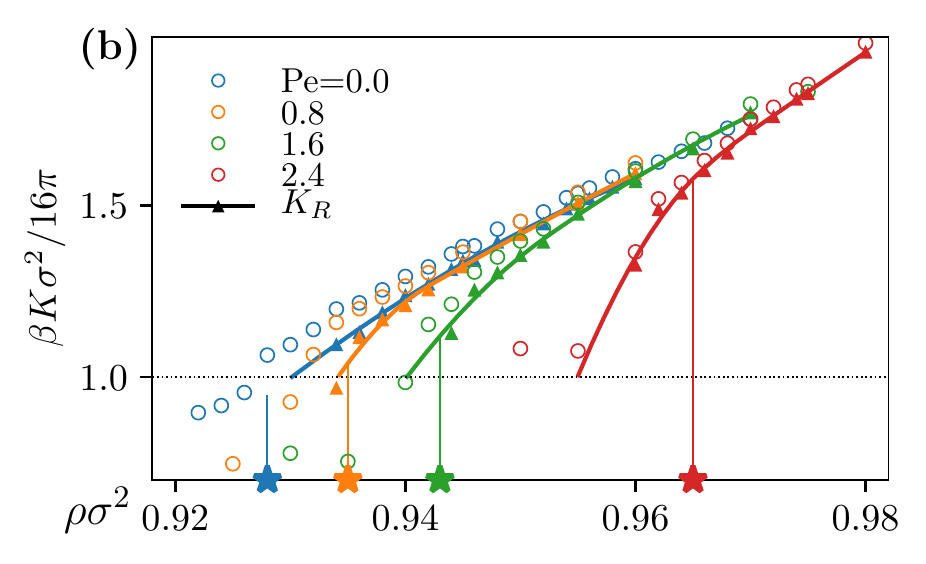}	
	\caption{(a) Young's modulus $K$ as a function of density $\rho\sigma^2$ collapse onto a single master curve $K\propto\exp(a\rho^3+b\rho^2+c\rho+d)$ for $0.0\le\mathrm{Pe}\le71.5$. The plus ($+$) markers denote the discontinuous jump in  $K$ for the corresponding activity, cross ($\times$) markers show the densities where the defects disappear for a large system size of $N=72\times10^3$ particles, and star ($\star$) markers denote the transition densities from the decay of $\psi_T$, all offset vertically for clarity. (b) Bare Young's modulus $K$ (open circles) obtained directly from the Lam\'e coefficients and the corresponding renormalized values $K_R$ (triangles) for $\mathrm{Pe}=0.0, 0.8, 1.6$ and $2.4$. The vertical lines mark the hexatic-solid transitions identified from the decay of $\psi_T$ for $N=72\times10^3$ particles.}
	\label{fig:youngsmod}
	\end{figure}

In \autoref{fig:youngsmod}(a) we plot the Young's modulus $K = 4\mu(\lambda+\mu)/(\lambda+2\mu)$ obtained for $0.0\le\mathrm{Pe}\le71.5$ and identify the densities where $K$ shows a discontinuous transition. Interestingly, we observe that $K$ collapses onto a single master curve for all $\mathrm{Pe}$. This remarkable result can be explained from the fact that the swim contribution to the stress tensor is negligible in the solid, and hence the elastic moduli for active solids equals  the ones of their passive counterparts. Activity only shifts the stability region of the solid to higher densities. Additionally, in the bottom part of the figure we mark the densities where the defects disappear for larger systems of $N=72\times10^3$ particles as a cross ($\times$) as well as the hexatic-solid transition densities obtained from the decay of positional correlations as a star ($\star$) similar to the ones marked in \autoref{fig:deffract2}. We find that in the passive case the discontinuity in $K$ agrees well with the hexatic-solid transition. However, for $\mathrm{Pe}\ge7.2$ the discontinuity in $K$ agrees well with the (dis)appearance of defects.

In equilibrium systems, the KTHNY theory suggests a critical value of $16\pi$ for the renormalized Young's modulus $\beta K_R\sigma^2$ below which the solid is unstable to shear. The renormalization procedure corrects for the interactions of defects at finite temperature. We  apply the renormalization procedure to the elastic constants up to $\mathrm{Pe}\le2.4$ for which there is a finite fraction of defects at the hexatic-solid transition.

To evaluate the renormalized Young's modulus we first explicitly measure the probability of dislocation pairs $p_d$ in simulations and calculate the core energy $E_c$ of defects using $p_d=\exp(-2\beta E_c)Z(K)$ where $Z(K)$ is the `internal partition function' of a dislocation~\cite{Fisher1979,Sengupta2000}. In equilibrium, due to thermal fluctuations there is a finite probability for the formation of dislocation pairs and the dislocation energy $E_c$ near the melting transition is small but finite. As we increase the activity the concentration of dislocation defects near the melting transition reduces (see  \autoref{fig:deffract2}). This observation hints that the energy needed to create a dislocation pair becomes higher as we increase  activity. Conversely, we can interpret that the unbinding energy reduces with increasing activity which eases the dissociation of dislocation pairs into dislocations. We then apply the recursion relations of the KTHNY theory~\cite{Fisher1979,Sengupta2000} to obtain the renormalized Young's modulus $K_R$, shown in Fig.~\ref{fig:youngsmod}(b) as a function of density for Pe=0, 0.8, 1.6 and 2.4. In the same plot we also show the `bare' values $K$ as in Fig.~\ref{fig:youngsmod}(a). For $\mathrm{Pe}=0$, we find that the renormalized $K_R$ differs significantly 
from the `bare' value. The density at which $K_R$ drops to zero agrees well with our estimate of the hexatic-solid transition. 
However, as we increase the activity upto $\mathrm{Pe}=2.4$ we find that this is no longer valid. Hence, even for a small Pe we can already see that the predictions of KTHNY theory based on the elastic constants deviate significantly from the transitions as obtained from the  positional correlations of the particles. At higher activity there is a complete absence of defects at the hexatic-solid transition point and the transition is  entirely driven by activity.

\emph{Conclusions:\label{sec:conclusions}}
We found that  at high activity the 2D melting of  active Brownian particle solids into a hexatic state is not driven by an unbinding of dislocation pairs in contrast to  passive systems. The hexatic state at high densities is completely devoid of any defects, and this defect-free region widens with activity. The solid-hexatic transition might be driven by a growing length scale of regions of cooperative motion, but this requires further investigation. Interestingly, we observed that the elastic constants of active solids  are equal to those of the passive counterparts, as the swim contribution to the stress tensor is negligible in the solid state. The activity only shifts the  stability regime of the solid state to higher densities. 

\begin{acknowledgments}
We thank Berend van der Meer, Laura Filion and Frank Smallenburg  for many useful discussions. S.P. and M.D. acknowledge funding from the Industrial Partnership Programme ``Computational Sciences for Energy Research'' (Grant No.14CSER020) of the Foundation for Fundamental Research on Matter (FOM), which is part of the Netherlands Organization for Scientific Research (NWO). This research programme is co-financed by Shell Global Solutions International B.V.
\end{acknowledgments}

%

\pagebreak
\widetext
\begin{center}
 \textbf{\large Supplementary Information: \\
 The role of topological defects in the two-stage melting and \\ elastic behavior of active Brownian particles}
 \date{\today}
\end{center}

\setcounter{equation}{0}
\setcounter{figure}{0}
\setcounter{table}{0}
\setcounter{page}{1}
\makeatletter
\renewcommand{\theequation}{S\arabic{equation}}
\renewcommand{\thesection}{S\arabic{section}}
\renewcommand{\thefigure}{S\arabic{figure}}

\section{Model}
We consider a two-dimensional system of isotropic Brownian particles that exhibit a self-propulsion speed $v_0$ which is directed along the orientation vector $\mathbf{e}_i=(\cos\theta_i,\sin\theta_i)$ assigned to particle $i$. To describe the translational and rotational motion of the individual colloidal particle $i=1,\dots,N$ we employ the overdamped Langevin dynamics:
	\begin{align}
		\gamma\dot{{\mathbf{r}}}_i &= -\sum_{j \neq i}^{} \boldsymbol{\nabla}_i \mathrm{U} (\mathrm{r}_{ij})
		+ \gamma v_0\mathbf{e}_i
		+ \sqrt{2\gamma k_BT} \boldsymbol{\Lambda}_i^{t},
		\nonumber
		\\
		\dot{\theta}_i &= \sqrt{2D_r} \Lambda_i^r, 
		\label{eqn:rotation}
	\end{align}
where $\gamma$ is the damping coefficient due to the drag forces from the implicit solvent, $k_B$ is the Boltzmann constant, and $T$ is the bath temperature. $D_r$ is the rotational diffusion coefficient. The quantities $\boldsymbol{\Lambda}_i^{t}$ and $\Lambda_i^{r}$ are unit-variance Gaussian noise terms with zero mean:
	\begin{align}
		\left\langle \boldsymbol{\Lambda}_i^{t}(t) \right\rangle  = 0, \quad&\quad \left\langle \Lambda_i^{r}(t) \right\rangle = 0,\nonumber
		\\
	\left\langle \boldsymbol{\Lambda}_i^{t}(t) \boldsymbol{\Lambda}_j^{t}(t') \right\rangle &= \mathbb{I}_2\delta_{ij} \delta(t-t') 
		\nonumber \\
	\left\langle \Lambda_i^{r}(t) \Lambda_j^{r}(t') \right\rangle
		&= \delta_{ij}\delta(t-t'),
	\end{align}
where $\mathbb{I}_2$ is the $2\times2$ identity matrix. The angular brackets $\langle\cdots\rangle$ denote an average over different realizations of the noise. The particles interact with a short-range repulsive Weeks-Chandler-Andersen (WCA) potential given by:
	\begin{align}
		\mathrm{U} (\mathrm{r}) &= 4\varepsilon\left[	\left(\frac{\sigma}{\mathrm{r}}\right)^{12} 
		-\left(\frac{\sigma}{\mathrm{r}}\right)^6 \right] + \varepsilon, 
					&& \mathrm{r} \leq 2^{1/6}\sigma	\nonumber \\
			&= 0 	&& \mathrm{r} > 2^{1/6}\sigma
	\label{eqn:wca}
	\end{align}
where $\mathrm{r} = |\mathbf{r}_{ij}|$ is the distance between the centers of particle $i$ and $j$, $\sigma$ is the particle diameter pertaining to the length scale in WCA potential, and $\varepsilon$ is the strength of the particle interactions. 

We set the system temperature $k_BT/\varepsilon=1$, following Ref.~\onlinecite{Redner2013}, and the damping coefficient $\gamma\sigma^2/\varepsilon=1$ fixing the translational diffusion coefficient to correspond to the free diffusion of particles given by the Stokes-Einstein relation $D_{t} = \gamma^{-1}k_BT$. This sets our time scale as $\tau=\gamma\sigma^2/k_BT$. The rotational diffusion coefficient is set to $D_r\tau=3$ and we use a time step size $dt=10^{-5}\tau$ for numerically integrating the equations of motion. We define a non-dimensional P\'eclet number  $\mathrm{Pe}=v_0\tau/\sigma$ as the ratio of the persistence length of motion to the particle diameter and perform simulations using the HOOMD-Blue \cite{anderson2008,Glaser2015} package in the range $0\le\text{Pe}\le150$. We used $N=72\times10^3$ particles in an approximately square 2D periodic simulation box with dimensions $L_x,L_y\approx 250\sigma$ for identifying the fluid-hexatic-solid transitions and $N=2.8\times10^3$ for calculating the elastic moduli. We used a regular hexagonally packed arrangement of particles at the overall system density as our initial configuration and we collect snapshots for $200\tau-500\tau$ at an interval of $\tau$ for analysis after allowing the system to achieve a stationary state for about $200\tau$.

As mentioned in the main text, the state diagram shown in Fig.~1 was obtained by measuring the equation of state (pressure-density curves), the density histograms and the decay of the orientational and the positional correlation functions to locate the boundaries as precisely as possible. Specifically, to locate the coexistence, we identified negative slope regions in the pressure-density curves as well as double-peaked structure of the density histograms similar to the analysis presented in Ref.~\onlinecite{Digregorio2018} and Ref.~\onlinecite{Cugliandolo2016}. We describe below the orientational and positional order parameters used for identifying the hexatic and solid phases.

\subsection{Orientational and Positional order}
\label{ch06:sec:transorder}
We measure the local 6-fold orientational symmetry around particle $i$ using the hexatic order parameter $\psi_6({\mathbf{r}}_i)$ given by:
\begin{equation}
\psi_{6}({\mathbf{r}}_i) = \frac{1}{N_b} \sum_{j\in N_b}^{} \exp(\iota 6\theta_{ij}),
\end{equation}
where $N_b$ denotes the number of nearest neighbors of  particle $i$ and the bond angle $\theta_{ij}$ is measured as a deviation of the orientation of the vector $\mathbf{r}_{ij}$ from the reference  global system orientation measured from $\Psi_6(L)$ averaged over all the particles. We identify the nearest neighbors $N_b$ of the particle by using a Voronoi construction.

\begin{figure}[h]
	\centering	
	\includegraphics[width=0.80\columnwidth]{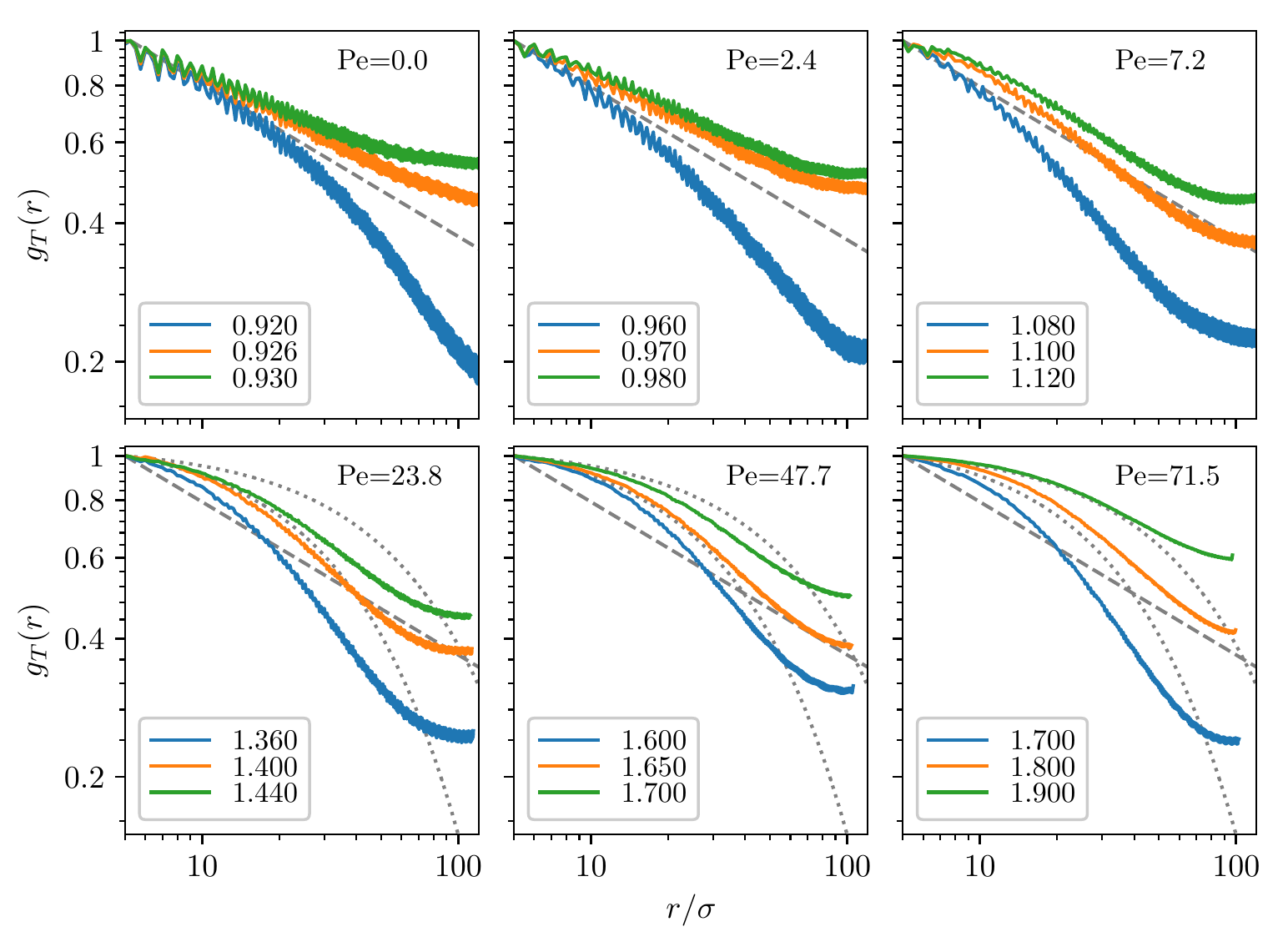}
	\caption{Positional correlation function $g_T(r)$ for $0\le\mathrm{Pe}\le71.5$ at the labeled densities. The decay is exponential for a hexatic state with the correlation length diverging upon increasing the density and the decay becomes quasi-long ranged for a crystalline state. The dashed grey line indicates algebraic decay with exponent $\eta_T=1/3$ and the dotted grey lines indicate exponential decay with correlation lengths $50\sigma$ and $100\sigma$.}
	\label{ch06:fig:gtrans}	
	\end{figure}

To investigate the decay of positional order, we  measure  the positional correlation function
\begin{equation}
		g_T(\text{r}) = \langle \psi_T^{*}(\mathbf{r}'+\mathbf{r})\psi_T(\mathbf{r'}) \rangle,
\end{equation}
where $\psi_T({\mathbf{r}}_i)$ is the positional order parameter expressed as:
		\begin{equation}
		\psi_T({\mathbf{r}}_i) = \exp(\iota \mathbf{k}_0\cdot{\mathbf{r}}_i).
		\end{equation}
Here $\mathbf{k}_0$ is the vector in reciprocal space denoting one of the first Bragg peaks in the 2D structure factor $S(\mathbf{k})$. The magnitude of this vector is equal to that of the reciprocal lattice vector i.e. $\mathbf{k}_0=(0, 4\pi/a\sqrt{3})$ where $a=(2/\rho\sqrt{3})^{1/2}\sigma$ is the lattice spacing in a regular hexagonal packing at a number density $\rho$ in a 2D geometry. According to the KTHNY theory, the positional order of a two-dimensional solid decays algebraically as $g_T(r)\propto r^{-\eta_T}$ with an exponent $0 \leq \eta_T \leq 1/3$.  Upon melting, the decay of the positional correlations becomes exponential i.e. $g_T(r)\propto \exp(-r/\xi_T)$ with  a correlation length $\xi_T$, which decreases with decreasing density. We show the positional correlation functions $g_T(r)$ as a function of particle separation $r$ in \autoref{ch06:fig:gtrans} and extract the correlation lengths $\xi_T$ in the case of an exponential decay or the exponent $\eta_T$ in the case of an algebraic decay. We identify the hexatic-solid transition by locating the density at which the exponent $\eta_T$ becomes smaller than $1/3$. For $\mathrm{Pe}=0$, we find that the decay of $g_T(r)$ becomes algebraic with $\eta_T\approx 1/3$ at $\rho\sigma^2=0.926$, marking the hexatic-solid phase transition. We locate the transition densities for higher Pe in a similar manner.

\subsubsection{System-size dependence}
\begin{figure*}[h]
	\centering
	\includegraphics[width=0.75\textwidth]{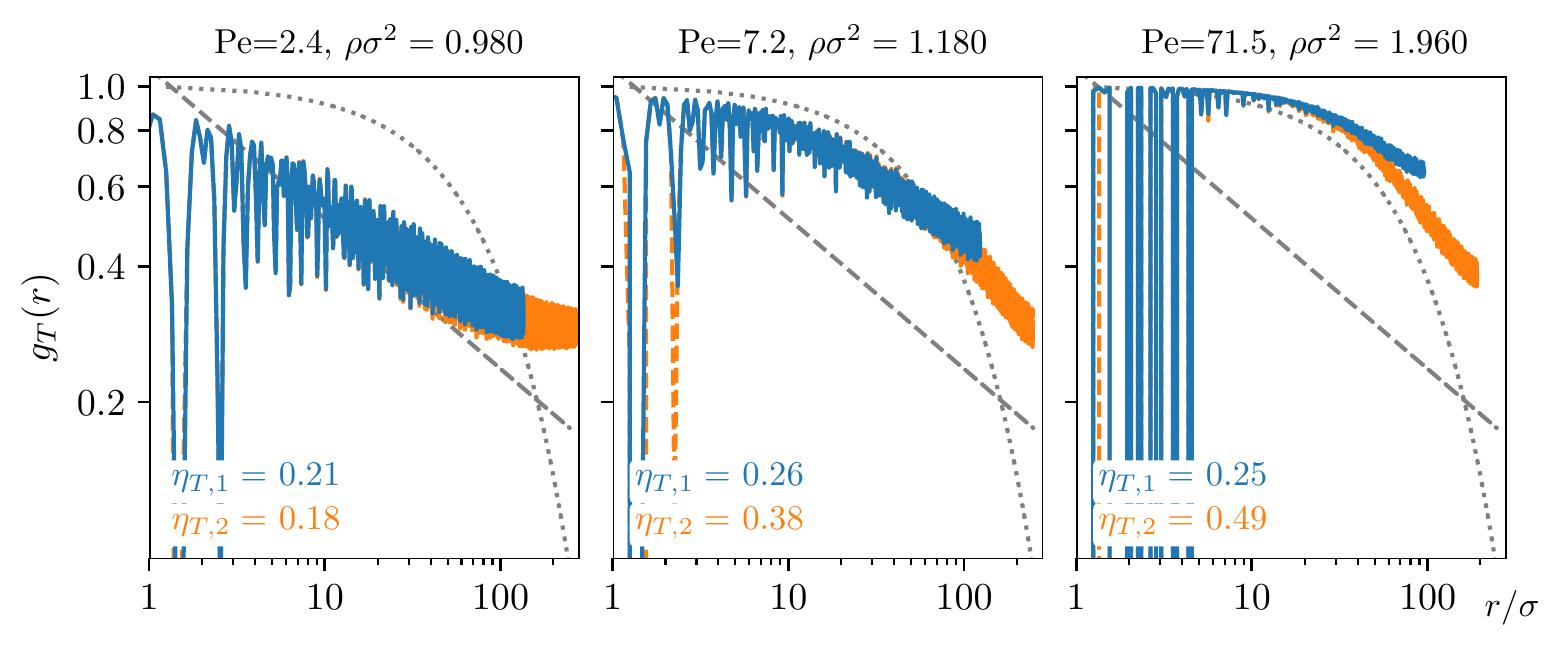}
	\caption{Comparison of the decay of positional correlations $g_T(r)$ on a $\log-\log$ scale for system sizes of $N=72\times10^3$ (orange lines) and $288\times10^3$ (blue lines) for Pe=$2.4, 7.2$ and $71.5$ at varying densities near the hexatic-solid transition. The exponents $\eta_{T,1}$ and $\eta_{T,2}$, obtained by fitting $g_T(r)\propto r^{-\eta_T}$ in the range $30\sigma-80\sigma$ and $50\sigma-150\sigma$, for a small and large system, respectively, are also quoted in the figure. The grey dotted line indicates an exponential decay with a correlation length $\xi_T=100\sigma$, and the grey dashed line indicates a power-law decay with exponent $\eta_T=1/3$.}
	\label{ch06:fig:gTsystemsize}
	\end{figure*}

To check the finite-size effects on the decay of the positional order we also simulate a few cases with $N=288 \times 10^3$ particles. We show $g_T(r)$ for the two different system sizes at $\mathrm{Pe}=2.4, 7.2$ and $71.5$ and with densities near the respective hexatic-solid transition in \autoref{ch06:fig:gTsystemsize}. The values for the exponents $\eta_{T,1}$ and $\eta_{T,2}$ of the power-law decay, for the small and large systems, respectively, are also listed in \autoref{ch06:fig:gTsystemsize}. For smaller system sizes, we observe clearly that $\eta_{T,1}<1/3$ for all three $\mathrm{Pe}$ values, which corresponds  to the solid phase.  For $N=288 \times 10^3$, we find that the exponent $\eta_{T,2}$ is still close to $\eta_{T,1}$ for $\mathrm{Pe}=2.4$, but for higher Pe the exponent $\eta_{T,2}$ is significantly larger than $\eta_{T,1}$. Despite these differences in the decay of the positional correlations, the location of the density where the defects disappear does not change and the observed hexatic region devoid of defects is robust over these investigated system sizes.

	\begin{figure*}
	\centering
		\includegraphics[width=\textwidth]{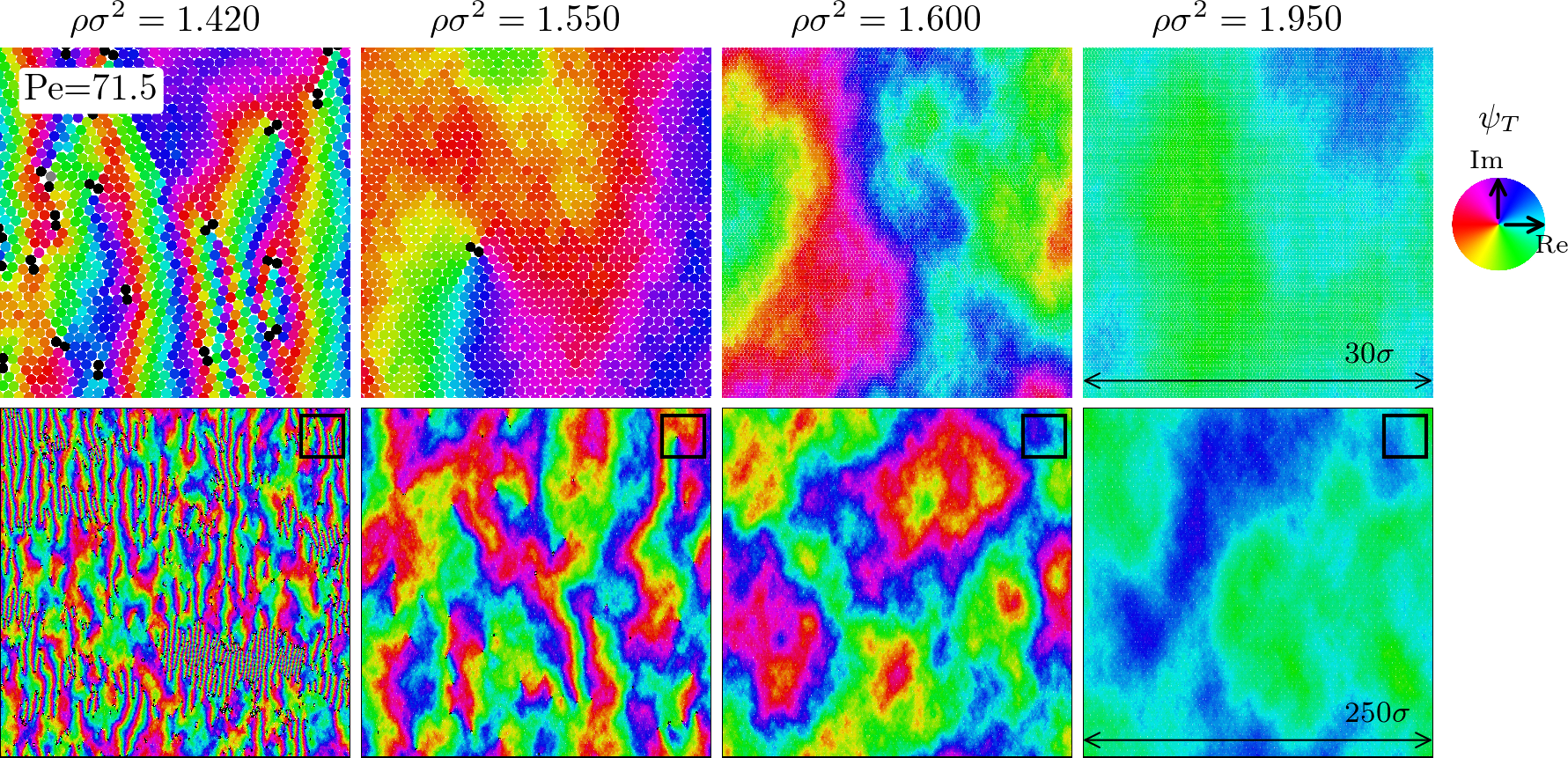}
	\caption{Typical particle configurations with the top row showing a magnification of the highlighted region shown in the bottom row by a black square for selected densities belonging to a pure hexatic phase ($\rho\sigma^2\le1.94$) and a solid phase ($\rho\sigma^2\ge1.95$) for $\mathrm{Pe}=71.5$. The color coding of the particles is according to $\arg(\psi_T)$ after subtracting the mean orientation, as shown in the color wheel on the right. We observe topological defects (colored black) in the configurations for $\rho\sigma^2=1.42$ and $1.55$ but not for $\rho\sigma^2=1.60$ and $1.95$ with $N=72\times10^3$ particles.
	\label{ch06:fig:snap_pe090}}
	\end{figure*}

\section{Elastic moduli}
\subsection{Stress tensor}
\label{ch06:sec:eos}
In order to  measure the bulk pressure $P$ in our system consisting of $N$ active Brownian particles, we employ the expressions as introduced by Winkler et al.\cite{winkler2015virial}, but modified them to the  2D case. Specifically, the pressure for isotropic active particles in a periodic box  with lateral dimensions $L_x$ and $L_y$ and 2D `volume' $V=L_x L_y$ is calculated using $P=\mathrm{Tr}(\mathbb{P})$ where the full stress tensor $\mathbb{P}$ is given by:
	\begin{equation}
		P_{\alpha\beta} = P_{\alpha\beta}^{\text{vir}} + \delta_{\alpha\beta}(P_{\alpha\beta}^{\text{id}} + P_{\alpha\beta}^{\text{swim}}).
		\label{ch06:eqn:fullP}
	\end{equation}	
Here $P^\mathrm{id}$ is the ideal gas pressure given by $P^\mathrm{id}=\rho k_BT$ with $\rho=N/V$ the number density of the particles. The virial contribution $P^\mathrm{vir}$ is obtained using the standard virial expression
	\begin{equation}
		P^\mathrm{vir}_{\alpha\beta} = -\frac{1}{4V}\left< \sum_{i}^{N}\sum_{j\neq i}^{N} \boldsymbol{\partial}_{\mathbf{r}_{i,\beta}} \mathrm{U} (r_{ij}) \cdot (\mathbf{r}_{i,\beta}-\mathbf{r}_{j,\beta})\right>.
	\end{equation}
The swim pressure contribution $P_\mathrm{swim}$ due to the self-propulsion is given by:
	\begin{equation}
		P^\mathrm{swim} = \frac{\gamma\rho v_0^2}{2D_r}
		-\frac{\gamma v_0}{4VD_r} \left< \sum_{i=1}^{N}\sum_{j\neq i}^{N} \boldsymbol{\nabla}_i \mathrm{U} (r_{ij}) \cdot \mathbf{e}_{i} \right>.
	\end{equation}

\subsection{Stiffness tensor and Lam\'e elastic coefficients}
In the linear elastic theory of isotropic solids, the elastic moduli relate the stress response of a system to an applied strain. In equilibrium, the elastic moduli are related to the free energy change due to such deformations~\cite{Landau1989theory}. Instead, for non-equilibrium systems we directly assume Hooke's law which linearly relates the mechanical stress $P_{\alpha\beta}$ with the applied strain $\epsilon_{\gamma\delta}$ through a symmetric stiffness tensor $\mathbb{C}$ given by:
	\[
		\mathbb{C} =	
		\begin{bmatrix}
		    C_{11}  & C_{12} & 0 & 0  \\
		    		& C_{22} & 0 & 0  \\
		    		& 		 & 0 & 0  \\ 
		    		& 		 & 	 & C_{44}
		\end{bmatrix}
		=
		\begin{bmatrix}
		 \lambda+2\mu   & \lambda		& 0	& 0  \\
			    		& \lambda+2\mu  & 0	& 0  \\
		    			& 		 		& 0	& 0  \\ 
			    		& 		 		&  	& \mu
		\end{bmatrix}
	\]
where $\lambda$ and $\mu$ are the Lam\'e coefficients in equilibrium systems~\cite{Landau1989theory}. For conciseness, we follow the Voigt notation above for indexing $C_{\alpha\beta\gamma\delta}$ with $xx$=1, $yy$=2, and $xy$=4. If a system is under a uniform isotropic pressure, the stiffness tensor $\mathbb{C}$ can be rewritten in terms of an effective stiffness tensor $\mathbb{B}$ as~\cite{Ray1988,Frenkelsmit2001}:
	\begin{equation}
	B_{\alpha\beta\gamma\delta} = C_{\alpha\beta\gamma\delta} - P(\delta_{\alpha\gamma}\delta_{\beta\delta}+
							\delta_{\alpha\delta}\delta_{\beta\gamma} -\delta_{\alpha\beta}\delta_{\gamma\delta} )
	\label{ch06:eqn:effectivestiffness}
	\end{equation}
	\begin{align}
	B_{11} = C_{11} - P, &\quad	B_{22} = C_{22} - P, \nonumber \\
	B_{12} = C_{12} + P, &\quad	B_{44} = C_{44} - P, \nonumber
	\end{align}
where $P=(P_{xx}+P_{yy})/2$ is the uniform pressure. The bulk modulus $E$, the shear modulus $G$ and the Young's modulus $K$ are related to the Lam\'e coefficients in 2D as:
	\begin{equation}
	E = \lambda + \mu, \quad G = \mu, \quad K = \frac{4\mu(\lambda+\mu)}{\lambda+2\mu} = \frac{4EG}{E+G}.
	\label{ch06:eqn:elastic}
	\end{equation}
Furthermore, from equilibrium statistical thermodynamics the isothermal compressibility $\kappa=1/E$, where $E$ is the bulk modulus, is expressed as:
\begin{equation}
 \frac{1}{E} = \kappa = -\frac{1}{V}\left(\frac{\partial V}{\partial P}\right)_T = \frac{1}{\rho}\left(\frac{\partial \rho}{\partial P}\right)_T,
\end{equation}
which can also be measured directly from the slope of $P-\rho$ curves. Once the uniform pressure of the system and the stiffness tensor (or the effective stiffness tensor $\mathbb{B}$) are known, we obtain the elastic moduli from $\lambda$ and $\mu$ using \autoref{ch06:eqn:elastic}. In our simulations we apply the method of box deformations to numerically evaluate the stiffness tensor for a system of interacting particles in an $NVT$ ensemble~\cite{Frenkelsmit2001,Clavier2017}. 

We extract the four non-zero elements of the stiffness tensor $\mathbb{C}$ by performing three kinds of deformations of the simulation box following Ref.~\onlinecite{Clavier2017}.  In the first kind of deformation, the box is elongated or compressed along the $x$-direction by a small factor $\epsilon_{xx}$ such that the particle coordinates in the $x$-direction become $x^{\prime}=x(1+\epsilon_{xx})$ and the box length also becomes $L_x^{\prime}=L_x(1+\epsilon_{xx})$. Similarly, the box can be elongated or compressed along the $y$-direction corresponding to imposing a small linear strain $\epsilon_{yy}$. Both these deformations correspond to a change in the overall density of the system but the magnitude is kept small in order to stay in the linear response regime. The third deformation is of a shearing type in which we change the shape of the box by keeping the volume constant. The angle between the $x$ and $y$ dimension box vectors, $\mathbf{a}$ and $\mathbf{b}$ respectively, is changed from $\pi/2$ to $\pi/2 - \tan^{-1}(\epsilon_{xy})$. The particle positions are then transformed as $(x,y)\rightarrow(x+y\epsilon_{xy},y)$.

	\begin{figure}[h]
	\centering
	\includegraphics[width=0.8\textwidth]{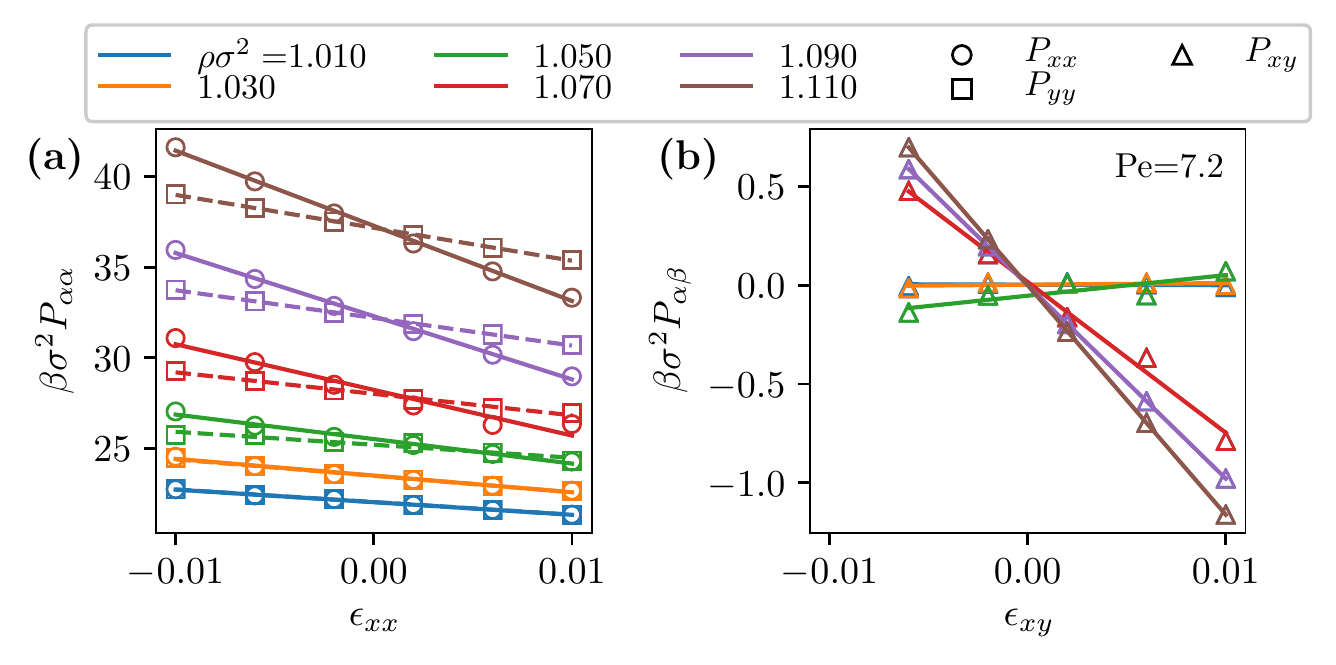}
	\caption{ (a) Diagonal $P_{xx} (\circ), P_{yy} (\square)$ and (b) off-diagonal $P_{xy} (\triangle)$ components of the full pressure tensor (\autoref{ch06:eqn:fullP}) obtained in a deformed simulation box with $N=2.8\times10^3$ particles as a function of linear tensile and shearing strains, $\epsilon_{xx}$ and $\epsilon_{xy}$, respectively, for various state points $\rho\sigma^2$ as labeled in the legend for $\mathrm{Pe}=7.2$. The stress response is linear in this regime of small strain magnitudes. We obtain the elements of the effective stiffness tensor $\mathbb{B}$ from the slope of a linear fit (solid and dashed lines) to the data points (symbols). The errorbars in the measurements are smaller than the symbol sizes.}
	\label{ch06:fig:stressstrain}
	\end{figure}

In our simulations, we start from a perfect hexagonal initial configuration with $N=2.8\times10^3$ particles and deform the box corresponding to the applied strain. We then measure the full stress tensor $P_{\alpha\beta}$ after a sufficiently long equilibration time that allows the system to reach a steady state. We perform the measurements by applying fixed linear strain $\epsilon_{xx}\in[-0.01,0.01]$ in intervals of 0.004. For an isotropic solid only the first two elements $C_{11}$ and $C_{12}$ are sufficient to obtain the Lam\'e coefficients $\lambda$ and $\mu$, which can be measured just by applying a longitudinal strain $\epsilon_{xx}$. However, for some cases we also measure the values of $\mu$ obtained by imposing a shearing strain $\epsilon_{xy}$ and confirm that the two independent measurements agree. The effective stiffness tensor $\mathbb{B}$ is directly obtained from the slope of a linear fit to the stress vs. strain curves, as shown for $B_{11}, B_{12}$ and $B_{44}$ in \autoref{ch06:fig:stressstrain}, using
	\begin{equation}
	B_{11} = \frac{\partial P_{xx}}{\partial \epsilon_{xx}}, \quad
	B_{22} = \frac{\partial P_{yy}}{\partial \epsilon_{yy}}, \quad 
	B_{12} = \frac{\partial P_{yy}}{\partial \epsilon_{xx}}, \quad
	B_{44} = \frac{\partial P_{xy}}{\partial \epsilon_{xy}}, \nonumber
	\end{equation}

\subsection{Bulk and Shear elastic moduli}
\begin{figure}[h]
	\centering
	\includegraphics[width=0.60\columnwidth]{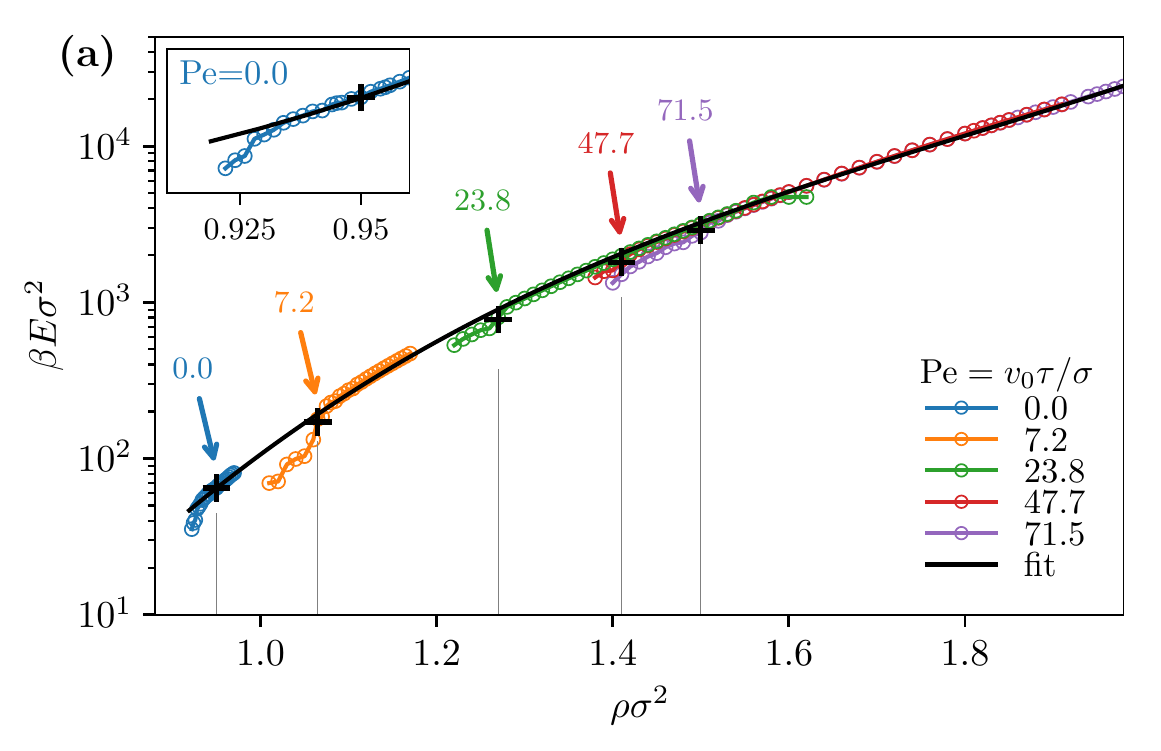}	
	\includegraphics[width=0.60\columnwidth]{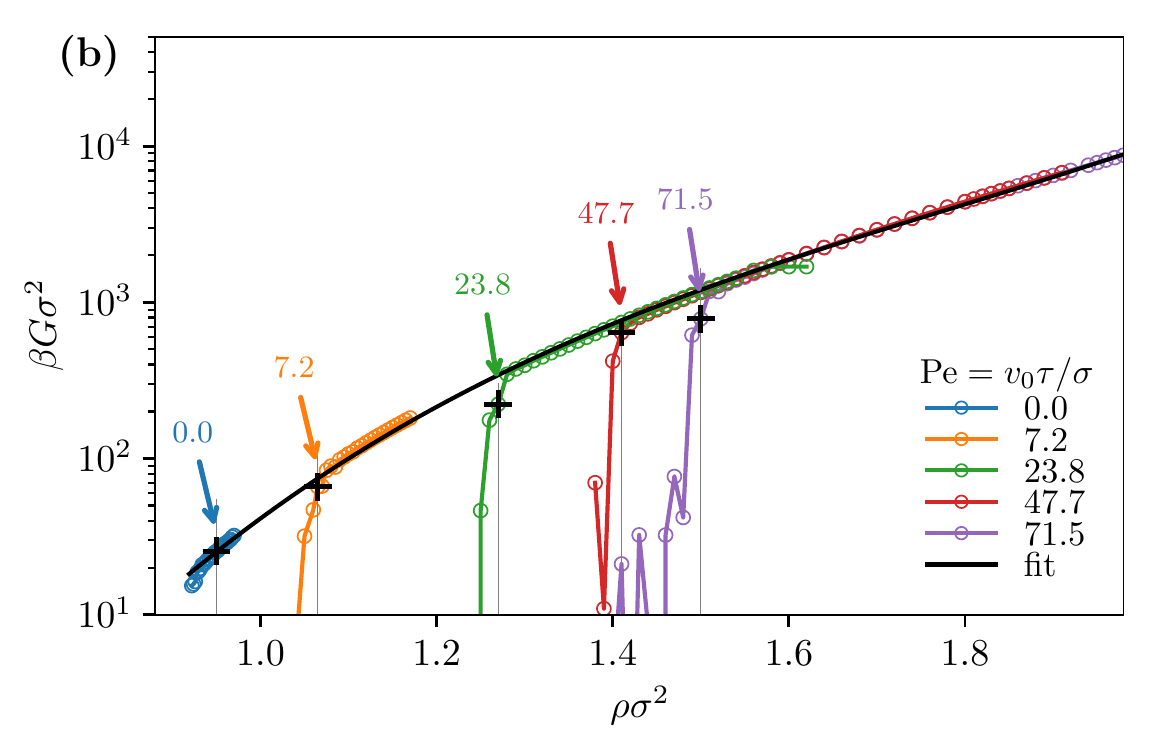}
	\caption{(a) Bulk modulus $E$ and (b) shear modulus $G$, obtained by explicitly straining the simulation box with $N=2.8\times10^3$ particles, as a function of density $\rho\sigma^2$ for various Pe as labeled in the legend. Both the bulk and the shear moduli collapse onto a single master curve indicated by the black lines which are fits of the form $E, G \propto\exp(a\rho^3+b\rho^2+c\rho+d)$. Upon lowering the density, the shear modulus $G$ drops sharply to zero at the critical density where the defects start to appear for the corresponding activity, and the bulk modulus $E$ shows a transition to a lower stable curve. The transition points at which the defect concentration becomes zero obtained from visual inspection in a system of $N=2.8\times10^3$ particles are marked with a plus ($+$).}
	\label{fig:figS8}
	\end{figure}

In \autoref{fig:figS8}(a) and \ref{fig:figS8}(b) we plot the bulk modulus $E$ and the shear modulus $G$, respectively, as a function of density for various Pe obtained using the method described above. For $\mathrm{Pe}=0$ (magnified in the inset) we find that there is a distinct jump in both $E$ and $G$, as indicated in the figure by a blue arrow, at a density of $\rho\sigma^2=0.926$. This jump is indicative of the second order nature of the transition. Upon increasing Pe, we observe a similar jump appearing in both $E$ and $G$ at higher densities marked by arrows in the figure. The bulk modulus $E$ shows only a discontinuity for higher Pe but the shear modulus $G$ shows a sharp drop to very small values at this transition upon reducing the density. Such a small value of the shear modulus $G$ indicates that the system is not a solid anymore and undergoes plastic deformation upon shearing. Furthermore, in the same plots we also indicate the densities where we observe a finite number of defects in the simulations with $N=2.8\times10^3$ particles by a plus marker ($+$) as in the main text Fig.~4(a). These points were determined by analyzing the sampled snapshots within our simulated time which show a complete absence of defects at densities higher than the marked points ($+$).

For $\mathrm{Pe}=0$ the defects disappear at a density of $\rho\sigma^2=0.950$ which is much higher than the point $\rho\sigma^2=0.926$ at which we observe the jump in the elastic moduli. For $\mathrm{Pe}\ge7.2$ we find that the two transition points agree extremely well. This indicates that for active cases the system becomes plastic as soon as a finite number of defects, mainly dislocations, appear in the system. On the other hand, the elastic moduli of the active solid states as a function of density collapse onto a single master curve independent of Pe. This remarkable result can be  explained by the fact that the swim contribution to  the stress tensor is zero or negligible in the solid phase, and hence, the elastic constants of active  solids become equal to those of passive solids at the same density. The sole effect of activity is that the  stable solid regime shifts  to higher densities  with activity. A numerical fit of the form $E, G \propto\exp(a\rho^3+b\rho^2+c\rho+d)$ is shown as a black solid line in both \autoref{fig:figS8}(a) and \autoref{fig:figS8}(b), and agrees very well with the measurements. 

\section{Renormalization procedure from KTHNY theory}
In equilibrium systems, the KTHNY theory suggests that the melting transition is accompanied by a lowering of the Young's modulus $\beta K$ below a critical value of $16\pi$. To correct for the interactions of defects present at a finite temperature a renormalization group analysis is applied to obtain the renormalized value $K_R$ of the Young's modulus which can then be compared against the numerical value of $16\pi$ to identify the melting transition.
The theory describes the dislocation defects in 2D systems associated with a `core energy' $E_c$~\cite{Kosterlitz1973,Nelson1979}. The probability of finding a bound pair of such dislocation defects is given by~\cite{Fisher1979,Sengupta2000}:
	\begin{align}
	p_d &= \exp(-2\beta E_c) Z(K) \nonumber \\
		&= \exp\left(-\frac{2E_c}{k_BT}\right) \frac{2\pi\sqrt{3}}{\beta K/8\pi-1} I_0\left(\frac{\beta K}{8\pi}\right)\exp\left(\frac{\beta K}{8\pi}\right)
		\label{ch06:eqn:prob_pairdisloc}
	\end{align}
where $Z(K)$ is the internal partition function of a dislocation, and $I_0$ is a modified Bessel function. 
The theory suggests a continuous transition from the solid to the hexatic state for large core energies $E_c\ge2.8k_BT$ and a weakly to strongly first-order transition as $E_c$ approaches and becomes lower than a value of $2.8k_BT$~\cite{Strandburg1988}. Typically, for systems with hard-core interactions the value of $E_c$ near the solid-hexatic transition is $\sim 6k_BT$ as found in Ref.~\cite{Qi2014} for monolayers of hard spheres.

The renormalization group recursion relations for the Young's modulus $K$ are expressed as \cite{Kosterlitz1973,Nelson1979,Sengupta2000}:
	\begin{align}
	\frac{\partial}{\partial l}\left(\frac{8\pi}{\beta K(l)}\right) &= 24\pi^2 y^2 \exp\left(\frac{\beta K}{8\pi}\right) \left[0.5I_0\left(\frac{\beta K}{8\pi}\right)-0.25I_1\left(\frac{\beta K}{8\pi}\right)\right]
	\label{ch06:eqn:rec1}
	\\
	\frac{\partial y(l)}{\partial l} &= \left(2-\frac{\beta K}{8\pi}\right)y + 2\pi y^2 \exp\left(\frac{\beta K}{16\pi}\right) I_0\left(\frac{\beta K}{8\pi}\right).
	\label{ch06:eqn:rec2}
	\end{align}
where the fugacity $y$ of the dislocation-pair fluid is obtained from an estimate of the core energy $E_c$ as:
	\begin{equation}
	y = \exp\left(-\frac{E_c}{k_BT}\right).
	\end{equation}	
The differential equations \autoref{ch06:eqn:rec1}-\ref{ch06:eqn:rec2} can be solved recursively for $l=0\dots\infty$ by using the unrenormalized (`bare') values $K(0)=K$ and $y(0)=\exp(-E_c(K(0)))$ as the initial guesses for $l=0$ and utilizing a trapezoidal (or higher order scheme) for performing the integration. The renormalized values are obtained from the renormalization-flow diagram of $y$-vs-$1/K$ (Fig.~1 in Ref.~\cite{Sengupta2000}) for the separatrix and $K_R=K(\infty)$ when $y(\infty)=0$. Exactly at the transition, the renormalization-flow follows the separatrix and above ($T>T_m$) and below ($T<T_m$) the melting point goes to the end points $\infty$ and $0$, respectively.

\clearpage
%

\end{document}